%% file: paper.tex
\documentclass[journal]{IEEEtran}
\IEEEoverridecommandlockouts
\usepackage{cite}
\usepackage{amsmath,amssymb,amsfonts}
\usepackage{algorithmic}
\usepackage{graphicx}
\usepackage{subcaption}
\usepackage{textcomp}
\usepackage{nicefrac}
\usepackage{soul}
\usepackage{pifont}
\usepackage{xcolor}

\renewcommand{\wr}{\psi}

\begin{document}

\title{\huge Ultra Low-Power and Real-time ECG Classification Based on STDP and R-STDP Neural Networks for Wearable Devices}

\author{Alireza~Amirshahi~and~Matin~Hashemi%
	
	{\color{blue} 
		\begin{flushleft}
			\footnotesize 
			This article is published. Please cite as A. Amirshahi, M. Hashemi, ``ECG Classification Algorithm Based on STDP and R-STDP Neural Networks for Real-time Monitoring on Ultra Low-Power Personal Wearable Devices," IEEE Transactions on Biomedical Circuits and Systems (TBioCAS), 2019. doi: 10.1109/TBCAS.2019.2948920
	\end{flushleft} }  
	
	%\thanks{Manuscript received November 1, 2018}
	\thanks{The authors are with Learning and Intelligent Systems Laboratory, Department of Electrical Engineering, Sharif University of Technology, Tehran, Iran. Webpage: http://lis.ee.sharif.edu, E-mails: alireza.amirshahi@ee.sharif.edu, matin@sharif.edu (corresponding author).}}

\maketitle
\begin{abstract} 
This paper presents a novel ECG classification algorithm for real-time cardiac monitoring on ultra low-power wearable devices. 
The proposed solution is based on spiking neural networks which are the third generation of neural networks. 
In specific, we employ spike-timing dependent plasticity (STDP), and reward-modulated STDP (R-STDP), in which the model weights are trained according to the timings of spike signals, and reward or punishment signals. 
Experiments show that the proposed solution is suitable for real-time operation, achieves comparable accuracy with respect to previous  methods, and more importantly, its energy consumption is significantly smaller than previous neural network based solutions.
\end{abstract}


\input{Introduction}
\input{Methods}

\input{Experiment}

%%%%%%%%%%%%%%%%%%%%%%%%%%%%%%%%%%%%%%%%%%%%%%%%%%%%%%%%%%%%%%%%%%%%%%%%%%%%%%%%%%%%%%%%%%%%%%%%%%%%%%%%%%%%%%%%%%%%%%%%%%%%
%%%%%%%%%%%%%%%%%%%%%%%%%%%%%%%%%%%%%%%%%%%%%%%%%%%%%%%%%%%%%%%%%%%%%%%%%%%%%%%%%%%%%%%%%%%%%%%%%%%%%%%%%%%%%%%%%%%%%%%%%%%%

\section{Conclusion}
\label{sec:conc}

This manuscript proposed an ECG classification algorithm in which both the pattern extraction and the classification steps are implemented based on spiking neural networks. 
In addition, the learning rules are optimized to better fit the ECG classification domain. 
As a result, the achieved accuracy is comparable to previous methods while the energy consumption is significantly smaller. 
Hence, the proposed method is suitable for ultra low-power wearable ECG monitoring devices.  

%\bibliography{ecg_snn_collection}
%\bibliographystyle{IEEEtran}


\end{document}

%% file: Introduction.tex
\section{Introduction}
\label{Introduction}

%CVD
Cardiovascular diseases (CVDs) account for a large proportion of global deaths \cite{CVDwho}. 
Early detection, as in many other diseases, plays a very crucial role in the process of stopping or controlling the progression of CVDs. 
Cardiac arrhythmias are widely used as signs for many of these diseases. Since the arrhythmias are reflected into electrical activities of the heart, they can be detected by analyzing Electrocardiogram (ECG) signals.

Visual inspection of recorded ECG signals during a visit to the cardiologist is the traditional form of arrhythmia detection. 
However, due to their intermittent occurrence, specially in early stages of the problem, arrhythmias are difficult to detect from a short time window of the ECG signal. 
Therefore, continuous ECG monitoring is crucial in early detection of potential problems \cite{tbiocas_mi_atienza_2018}.

%Wearables
In recent years, personal wearable devices have been introduced as cost-effective solutions for continuous ECG monitoring in daily life. 
Previous works include ECG monitoring solutions that are low-power but only extract the R peak or the QRS complex and do not further process the ECG signal \cite{Bayasi2015, tbiocas_2015_asic_r, tbiocas_2018_asic_qrs, jbhi_dsp_qrs_atienza_2018}. 
Many others analyze the signal and perform ECG classification, but transmit the signal to a connected smartphone or a remote cloud server for performing the computations associated with ECG classification algorithms \cite{jbhi_cloud_2014, jbhi_cloud_2015}. Besides privacy concerns \cite{privacy2018}, employing such solutions for continuous cardiac monitoring is limited by the availability, speed and energy consumption of the wireless connection. 
Other solutions include offline analysis of recorded ECG signals \cite{GR_NN_2017,jbhi_dsp_cluster_2018}. 
Our approach, on the other hand, is to continuously monitor the ECG signal in real-time and on the wearable device itself.

%Compute+Power
In this approach, the ECG signal is continuously monitored without any connection to or any assistance from other systems. 
However, the main limiting factor is that wearable health monitoring devices should be small, and thus, may not utilize powerful processors or large batteries. 
Therefore, computational costs and energy consumption of ECG classification algorithms are important for real-time operation on small and low-power wearable devices \cite{tbiocas_mi_atienza_2018}.

Classical algorithms were mainly based on morphological features, such as the QRS complex, and signal processing techniques such as Fourier transform and wavelet transform \cite{Minami1999, Shyu2004, Lee2015, tbiocas_2017_asic_qrs_based}. 
Such features show significant variations among different individuals and under different conditions, and therefore, are not sufficient for accurate arrhythmia detection \cite{Hoekema_2001, Kiranyaz2016}. 
%
%2,3 ann and dnn
Many solutions have been proposed based on artificial neural networks, and especially in recent years, deep convolutional neural networks (CNN) and recurrent neural networks (RNN) \cite{Hu1997, Ince2009, Rajpurkar2017, Kachuee2018, Kiranyaz2016, saeed2019}. Neural network algorithms automatically extract the features from data, and hence, provide higher accuracy. This is because neural network based  feature extraction is inherently more resilient to variations among different ECG waveforms \cite{Kiranyaz2016}. 
However, the downsides of employing deep neural networks are high computational intensity and large power consumption. 

%4 snn
Spiking neural networks (SNN) are the third generation of neural networks and provide the opportunity for ultra low-power operation \cite{MAASS1997,SNN_Silicon,Qiao2016,Merolla668,intel_loihi_2018,zeroth}. 
In this type of neural networks, the information is processed based on propagation of spike signals through the network as well as the timing of the spikes. Every neuron consumes energy only when necessary, that is only when it sends or receives spikes. 
In recent years, neuromorphic processors have been introduced for low-power implementation of SNN-based algorithms. Examples include IBM TrueNorth \cite{Merolla668}, Intel Loihi \cite{intel_loihi_2018} and Qualcomm Zeroth \cite{zeroth}.

\begin{figure}[tp]
	\begin{center}
		\centerline{\includegraphics[width=1\columnwidth]{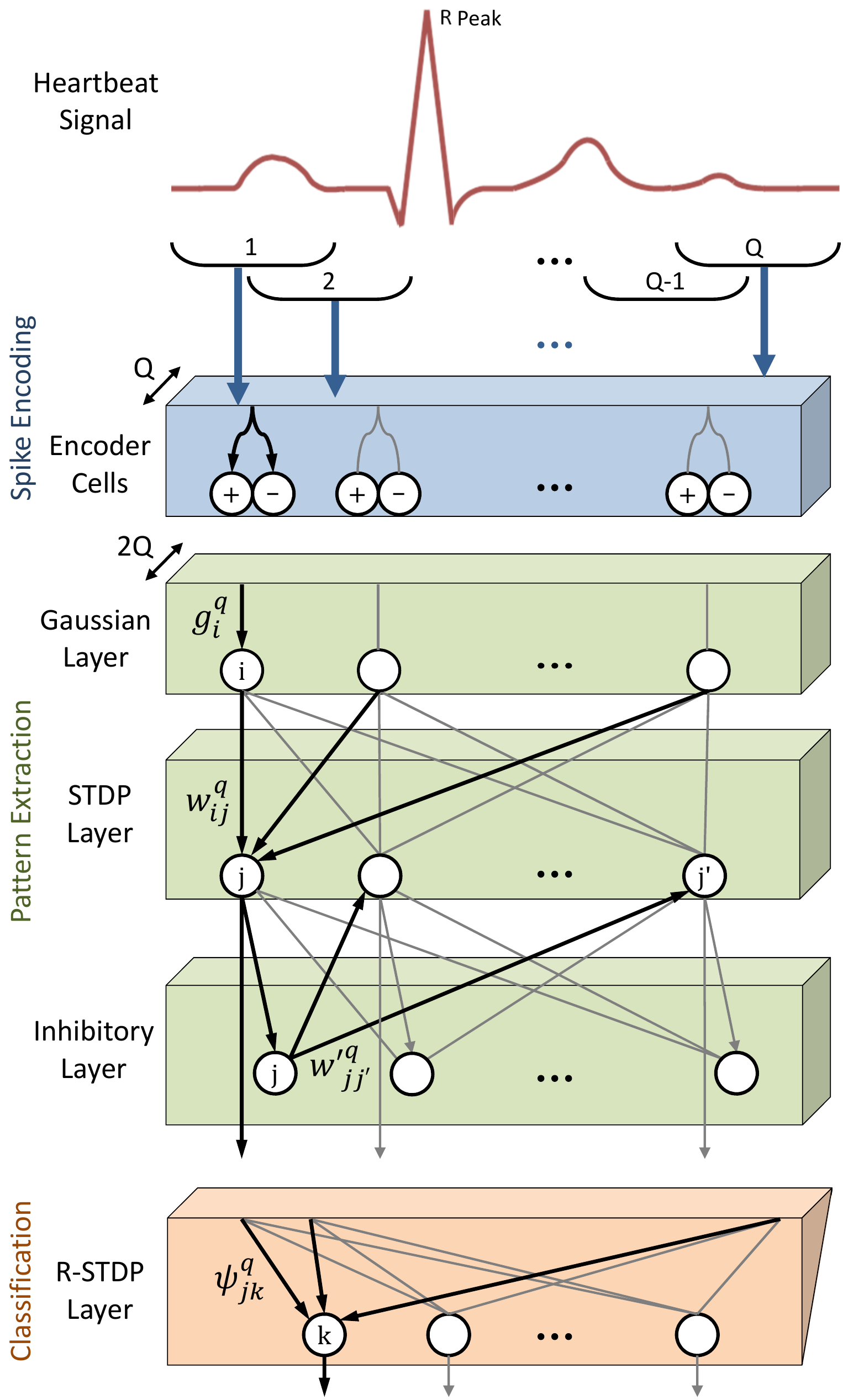}}
		\caption{Overall view of the proposed solution.}
		\label{fig:complete_model}
	\end{center}
\end{figure}

%our solution
This paper proposes a novel SNN-based ECG classification algorithm for real-time operation on ultra low-power wearable devices. The overall view of the proposed solution is shown in Fig.~\ref{fig:complete_model}. 
The heartbeat signal is split into a small number of overlapping windows, which are then encoded into spike signals. 
The patterns in the generated spikes are automatically extracted in the STDP layer with assistance from a Gaussian layer and an inhibitory layer. Finally, the extracted features are classified in the R-STDP layer. 
STDP stands for spike-timing dependent plasticity which means the timings of the spikes are used to train the weights in this layer. R-STDP stands for reward-modulated STDP which means reward or punishment signals are used to train the weights, in addition to the spike timings. 
All the above layers are performed in the spike domain, and therefore, energy consumption is reduced significantly. 
In addition, the learning rules in training STDP and inhibitory layers are optimized in order to better fit the ECG classification domain. 
Experiments show that the proposed solution is suitable for real-time operation, achieves comparable classification performance with respect to previous methods, and more importantly, its energy consumption is significantly smaller than previous neural network based methods.

Previous SNN-based solutions that process ECG signals employ non-SNN algorithms for pattern extraction \cite{Kolagasioglu2018}, or do not fully analyze the ECG signal, for instance, extract the heart rate \cite{Das2018}, apply a band-pass filter to the input ECG signal \cite{filter2013}, or detect perturbations in the signal \cite{Cattaneo2018}.

%other areas
It is noteworthy to mention that SNNs have previously been employed in many areas in computer vision. Examples include digit recognition~\cite{Diehl2015a, Iakymchuk2015} and visual categorization~\cite{Masquelier2007, Kheradpisheh2018, Mozafari2018}. This is mainly because the spiking neuron models are inspired from neuroscience studies on the brain's visual cortex. 
SNNs have also been applied to non-vision applications including robot path planning and control~\cite{Hwu2017,bing2018survey} and classification of EEG spatio-temporal data~\cite{Kasabov2015}.

The rest of this manuscript is organized as the following. Section~\ref{sec:snn} provides a brief introduction to the neuron model and other basic concepts related to spiking neural networks. Section~\ref{sec:alg} presents our proposed SNN-based ECG classification algorithm.  Section~\ref{sec:exp} presents the results of experimental evaluations as well as comparisons with previous works in term of classification performance, network type, and energy consumption. Section~\ref{sec:conc} concludes the paper.

%% file: Methods.tex
\section{Neuron Model}
\label{sec:snn}

\begin{figure}[tp]
	\begin{center}
		\centerline{\includegraphics[width=0.75\columnwidth]{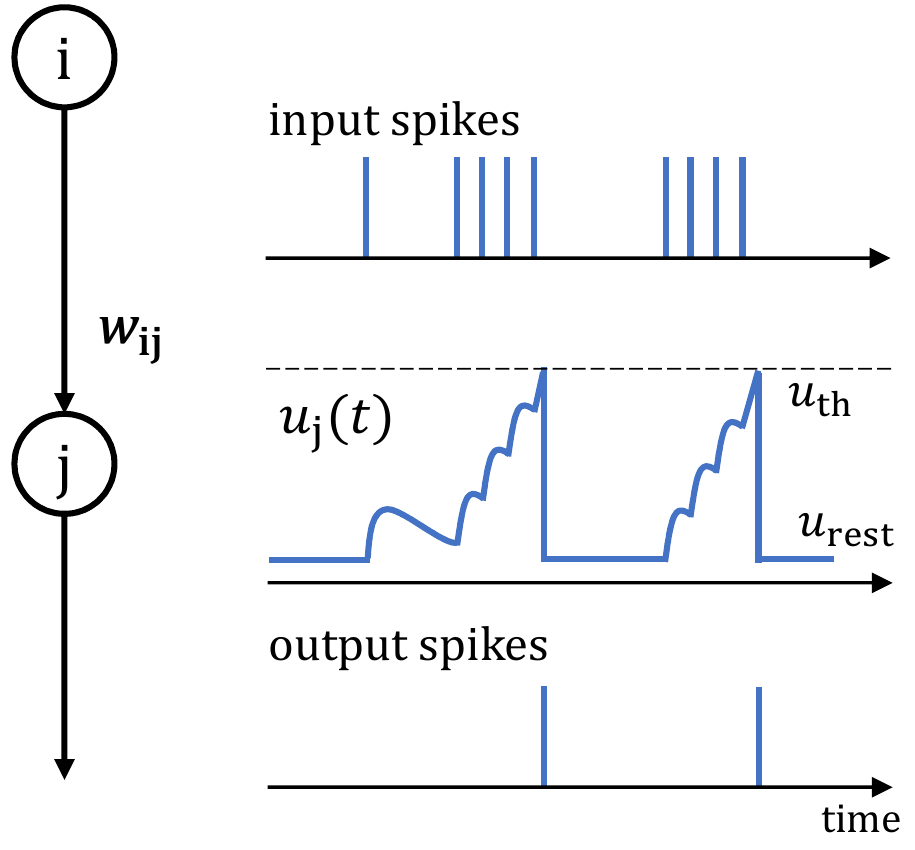}}
		\caption{In this example, neuron $i$ sends nine spikes to neuron $j$. As a result, the membrane potential of neuron $j$, i.e., $u_j$, is increased and reached to $u_{th}$ twice. Hence, two spikes are fired on its output.}
		\label{fig:pre_post}
	\end{center}
\end{figure}

The employed neuron model and other basic concepts in spiking neural networks (SNN) are briefly explained in this section. 
SNNs are the third generation of neural networks which are inspired by the nerve cells in the brain. In contrast to the second generation neural networks such as multi-layer perceptrons, here the concept of time is deeply incorporated into the operational model \cite{Gerstner2014}. 
In specific, the \emph{neurons} communicate through electrical signals called \emph{spikes}. For example in Fig.~\ref{fig:pre_post}, neuron $i$ sends nine spikes to neuron $j$ at different times. The information is encoded in the timing of the spikes, not in their amplitude. 

A neuron $i$ fires at time $t$, i.e., generates a spike on its output at time $t$, only when its \emph{membrane potential} $u_i(t)$ reaches a specific threshold value $u_{\text{th}}$. When neuron $i$ fires, $u_i$ is decreased to $u_{\text{rest}}$ and the generated spike travels to neuron $j$ and causes an increase  in $u_j$, i.e., the membrane potential of neuron $j$. This is shown in Fig.~\ref{fig:pre_post}. 
Since neurons do not fire all the time, the energy consumption is significantly smaller than other types of neural networks \cite{Merolla668}. 

A connection from neuron $i$ to $j$ is called a \emph{synapse}. For this synapse, $i$ is called \emph{pre-synaptic} neuron, and $j$ is called \emph{post-synaptic} neuron. A \emph{weight} $w_{ij}$ is associated with every synapse. 

Different brain-inspired mathematical models exist for SNNs. We employ one of the most popular models called Leaky Integrate-and-Fire~(LIF) model \cite{Gerstner2014} which operates based on the following  equation. This differential equation describes how $u_j$ is changed at time $t$. 
\begin{equation}
\label{eq:LIF}
 \tau \frac{d}{dt}u_j(t) = - \big( u_j(t) - u_{\text{rest}} \big) + \alpha \sum_i s_i(t)w_{ij}
\end{equation}

The index $i$ iterates through all neurons whose outputs are connected to neuron $j$. The term $s_i(t)$ is equal to either one or zero, and represents whether neuron $i$ has generated a spike at time $t$. 
Hence, the term $\sum_i s_i(t)w_{ij}$ is zero when no spike is received by neuron $j$ at time $t$. When one or multiple spikes are received at time $t$, the term $\sum_i s_i(t)w_{ij}$ causes an increase in $\frac{d}{dt}u_j(t)$. 
In addition, a leaky current causes membrane potential $u_j$ to tend to the rest value $u_{\text{rest}}$ with time constant $\tau$. 
To summarize, $u_j$ does not change at time $t$, if it is equal to its rest value $u_{\text{rest}}$ and no spike is received. Otherwise, $\frac{d}{dt}u_j(t)$ is non-zero, and thus, $u_j(t)$ changes at time $t$.

According to the above operational model, synaptic weight $w_{ij}$ impacts spike propagation from neuron $i$ to $j$. A larger $w_{ij}$ means a spike generated by neuron $i$ is more likely to trigger a spike by neuron $j$. This is because $s_i(t)w_{ij}$ affects $\frac{d}{dt}u_j(t)$. As a result, the larger the weight $w_{ij}$ the higher the firing rate of neuron $j$.

%%%%%%%%%%%%%%%%%%%%%%%%%%%%%%%%%%%%%%%%%%%%%%%%%%%%%%%%%%%%%%%%%%%%%%%%%%%%%%%%
%%%%%%%%%%%%%%%%%%%%%%%%%%%%%%%%%%%%%%%%%%%%%%%%%%%%%%%%%%%%%%%%%%%%%%%%%%%%%%%%
%%%%%%%%%%%%%%%%%%%%%%%%%%%%%%%%%%%%%%%%%%%%%%%%%%%%%%%%%%%%%%%%%%%%%%%%%%%%%%%%
\section{Proposed SNN-Based ECG Classification}
\label{sec:alg}

Our proposed SNN-based method for classification of ECG beats is presented in this section. Fig.~\ref{fig:complete_model} illustrates an overall view of the proposed solution. 
The input ECG signal is first segmented into heartbeats. The heartbeat is split into a small number of windows, which are then encoded into spikes. The  patterns in the generated spikes are automatically extracted in the STDP layer with assistance from two other layers, namely the Gaussian and the inhibitory layers. Finally, the extracted features are classified in the R-STDP layer. 
The formulations which govern the STDP and the inhibitory layers are optimized to better fit ECG classification problem. 
The details are discussed in the following. 

\subsection{Segmentation}
\label{sec:alg:seg}

Heartbeat segments are formed as $0.25$~seconds of the input ECG signal before an $R$ peak and $0.45$~seconds after. $R$ peak is a specific point in the ECG waveform as shown in Fig.~\ref{fig:complete_model}. There exist many ultra low-power and highly accurate methods for $R$ peak detection  \cite{tbiocas_2015_asic_r, tbiocas_2018_asic_qrs}. 

Since the $R$ peak has a much larger amplitude compared to other parts of the heartbeat, it becomes the dominant pattern in the STDP layer and basically causes the effect of other patterns to be highly reduced. In order to avoid this issue and efficiently capture all the patterns, every heartbeat is split into $Q$ overlapping windows as shown in Fig.~\ref{fig:complete_model}. 
The windows are processed separately in the pattern extraction step. Let's denote the heartbeat signal as $X_{ecg}$ and the portion of $X_{ecg}$ which falls in window $q \in [1,Q]$ as $X^q_{ecg}$. The number of samples in $X^q_{ecg}$ is given by the following equation. 
\begin{equation}
| X^q_{ecg} | = \frac{1}{\lceil Q/2 \rceil} \times | X_{ecg} |
\end{equation}

%%%%%%%%%%%%%%%%%%%%%%%%%%%%%%%%%%%%%%%%%%%%%%%%%%%%%%%%%%%%%%%%%%%%%%%%%%%%%%%%
\subsection{Spike Encoding}
\label{sec:methods:encoder}

There is an encoder cell for every sample $i \in [1,| X^q_{ecg} |]$ from every window $q \in [1,Q]$. Every cell encodes its input into a series of spikes as the following. 
Spike generation is random and follows the Poisson process. 
The spike firing rate of this random Poisson process is set proportional to $X^q_{ecg}[i]$, i.e., the amplitude of sample $i$ from window $q$, plus a small bias. The higher the amplitude, the higher the rate of spike firing of the corresponding cell  \cite{Dayan2001}. 

In fact, since the amplitude can be both positive or negative, two encoder cells are employed for every sample, not one. Spikes are generated by the first encoder cell if the signal is positive, and by the second cell if negative. This is inspired by the nerve cells in the Retina \cite{Meister1999}. 

The output of every encoder cell is connected to one synapse, and the generated spikes are fed into this synapse. 
Therefore, the total number of output synapses for a window $q \in [1,Q]$ is equal to 
\begin{equation}
2 \times | X^q_{ecg} |
\end{equation}

These synapses are split in two groups. As shown in Fig.\ref{fig:complete_model}, the number of windows in the next layers is equal to $2Q$. The positive encoder cells are connected to the synapses in the odd windows, and the negative encoder cells are connected to the synapses in the even windows.

%%%%%%%%%%%%%%%%%%%%%%%%%%%%%%%%%%%%%%%%%%%%%%%%%%%%%%%%%%%%%%%%%%%%%%%%%%%%%%%% 
\begin{figure*}[tp]
\centering
\begin{subfigure}[]{0.3 \textwidth}
\includegraphics[width=\linewidth]{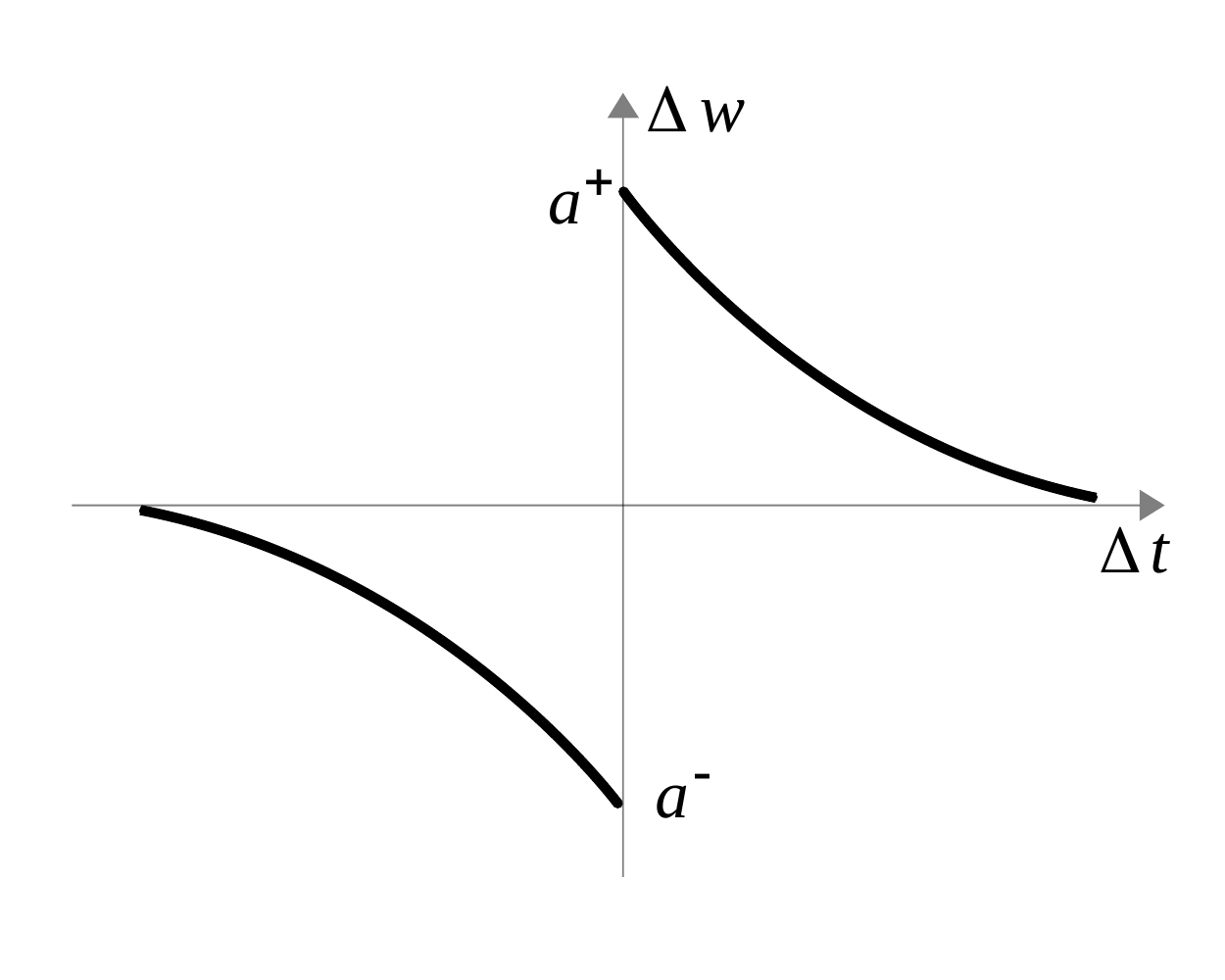}
\caption{}
\label{fig:usual_STDP}
\end{subfigure}
\begin{subfigure}[]{0.3 \textwidth}
\includegraphics[width=\linewidth]{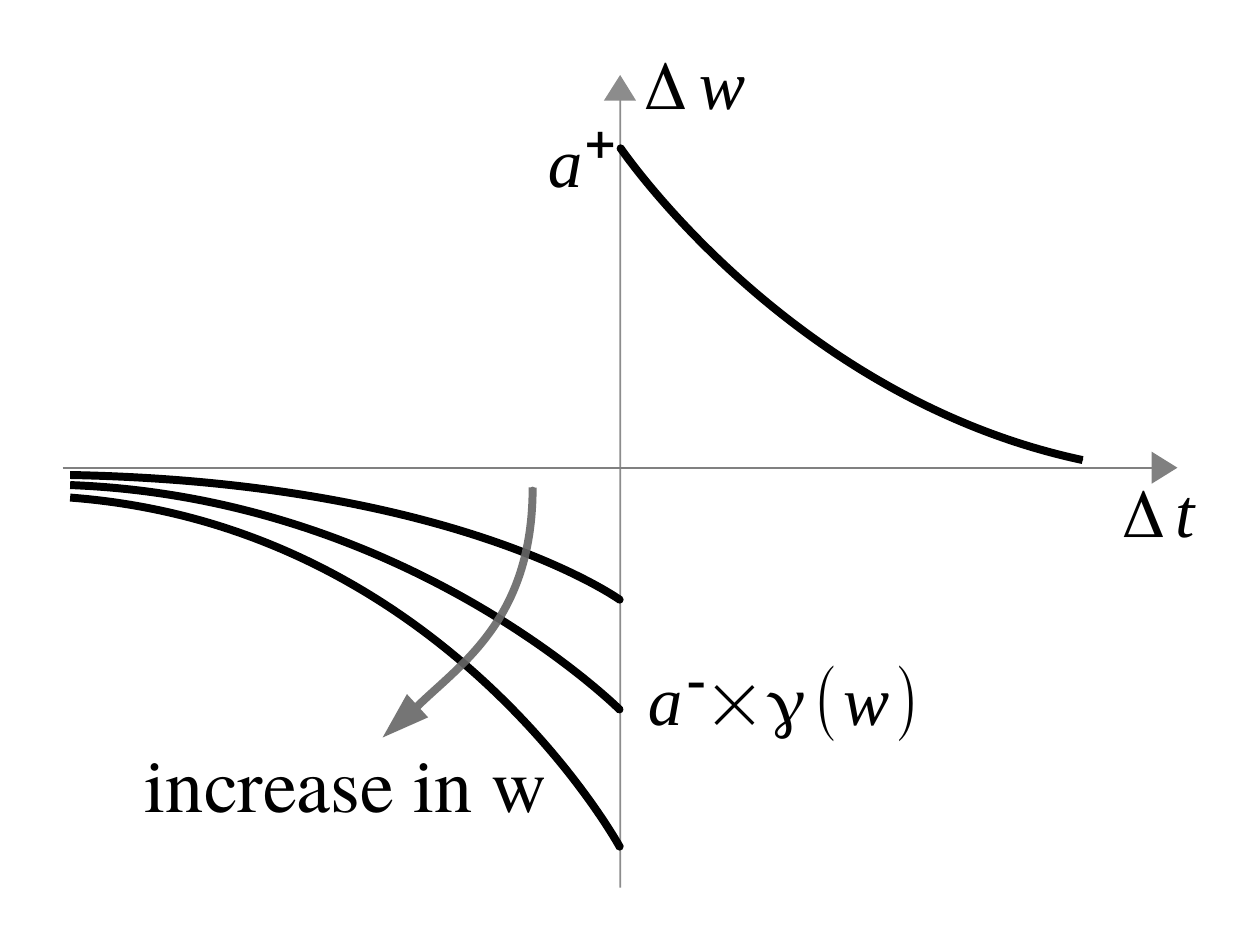}
\caption{}
\label{fig:our_STDP}
\end{subfigure}
\begin{subfigure}[]{0.3 \textwidth}
\includegraphics[width=\linewidth]{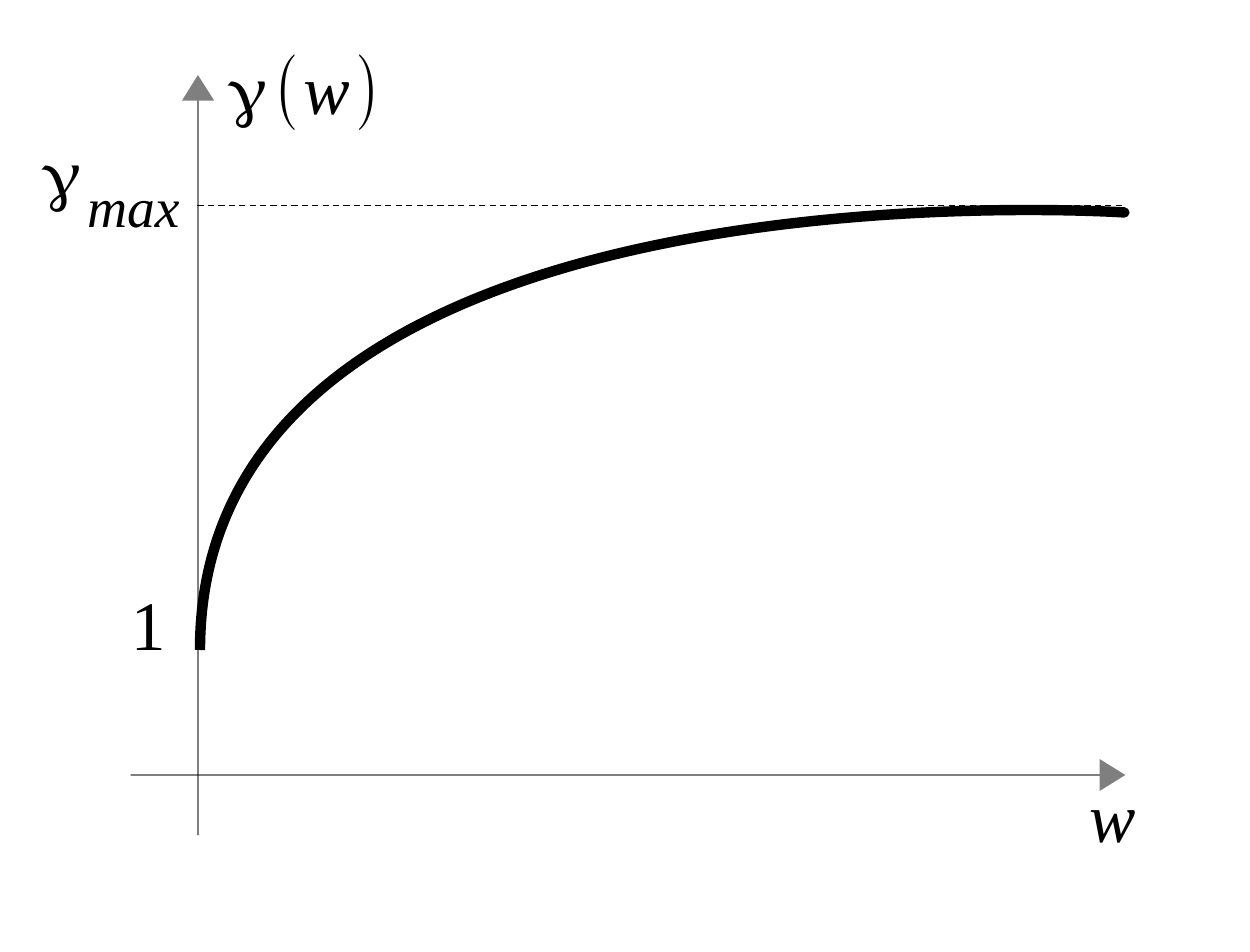}
 \caption{}
\label{fig:max_LTD}
\end{subfigure}
\caption{
Learning rule for the synaptic weights in the STDP layer. (a) The usual STDP learning rule. (b) Proposed learning rule. (c) The value of $\gamma(w)$ in (b) is between $1$ and $\gamma_{\text{max}}$.
}
\end{figure*}

%%%%%%%%%%%%%%%%%%%%%%%%%%%%%%%%%%%%%%%%%%%%%%%%%%%%%%%%%%%%%%%%%%%%%%%%%%%%%%%%
\subsection{Gaussian Layer}
\label{sec:alg:g}

The purpose of this layer is to adjust the number of spikes in order to help the STDP-based pattern extraction that is discussed later in Section \ref{sec:alg:stdp}. 

Every window $q \in [1,2Q]$ is processed separately as the following. 
As shown in Fig.\ref{fig:complete_model}, input synapse $i$ in window $q$ is connected to one neuron and its synaptic weight is set to $g_{i}^{q}$. As a result, the neuron adjusts the spike firing rate by a factor of $g_{i}^{q}$. In specific, we have 
\begin{equation}
\label{eq:gaussian_inout}
R^{q}_{\text{out}, i} = g_{i}^{q} \times  R^q_{\text{in}, i}
\end{equation}
where $R^{q}_{\text{in},i}$ and $R^{q}_{\text{out},i}$ denote the rate of spikes on the input and on the outputs of the neuron, respectively.

Since the windows overlap, an ECG peak which falls on the side of a window also appears in the middle of a neighbor window. In order to reduce the effect of side peaks, a Gaussian kernel is employed in setting the value of $g_{i}^{q}$. 
In specific, $g_{i}^{q}$ is set as 
\begin{equation}
\label{eq:gaussian_init}
g_{i}^{q} = 
\beta^q \times 
\frac{1}{\sigma\sqrt{2\pi}} \times
e^{
\displaystyle -\frac{1}{2} \big(\frac{i-\mu}{\sigma}\big)^2
}
\end{equation}
where $\mu=\frac{1}{2}| X^q_{ecg} |$ and $\sigma = \frac{1}{3}| X^q_{ecg} |$. Note that $| X^q_{ecg} |$ denotes the window length. The mean $\mu$ is chosen such that the Gaussian function be centered at the middle of the window. The variance $\sigma$ is chosen such that the effect of a side peak is reduced but not completely neglected.

The term $\beta^q$ is a trainable variable which is different for different values of $q$, i.e., different windows. It is initialized as $\beta^q=1$, and then trained as the following. 
The next layers operate more efficiently if all windows $q \in [1,2Q]$ in this layer generate equal average number of spikes. Let $\overline{R^q_{\text{out}}}$ denote the average number of spikes fired by the neurons in window $q$ in this layer. The larger the value of $\beta^q$, the larger the value of $\overline{R^q_{\text{out}}}$. 

Therefore, in training $\beta^q$ we would like to have $\overline{R^q_{\text{out}}}$ reach the same target rate $R_{\text{target}}$ for all $q\in[1,2Q]$. 
Since the neuron behavior (Section \ref{sec:snn}) is non-linear, to achieve the desired rate, $\beta^q$ is initialized to $1$, then, the train data is fed to the network, and $\beta^q$ is iteratively updated as the following. 
\begin{equation}
\label{eq:gaussian}
\Delta \beta^q = \alpha (1 - \frac{\overline{R^q_{\text{out}}}}{R_{\text{target}}}) 
\end{equation}

%%%%%%%%%%%%%%%%%%%%%%%%%%%%%%%%%%%%%%%%%%%%%%%%%%%%%%%%%%%%%%%%%%%%
%%%%%%%%%%%%%%%%%%%%%%%%%%%%%%%%%%%%%%%%%%%%%%%%%%%%%%%%%%%%%%%%%%%%%
%%%%%%%%%%%%%%%%%%%%%%%%%%%%%%%%%%%%%%%%%%%%%%%%%%%%%%%%%%%%%%%%%%%%%
\subsection{STDP-Based Pattern Extraction}
\label{sec:alg:stdp}

Spike-timing dependent plasticity~(STDP)~\cite{song2000} is employed here to extract the patterns in the ECG signal. STDP is an unsupervised learning procedure. Every window $q \in [1,2Q]$ is processed separately as the following. Let $w^q_{ij}$ denote the synaptic weights in window $q$ in this layer. 
The synaptic weights, also known as plasticity, are trained according to the timing patterns of the spikes in pre- and post-synaptic neurons. 
In specific, consider a synapse $ij$ in this layer which connects pre-synaptic neuron $i$ to post-synaptic neuron $j$. If the spike time of the pre-synaptic neuron, denoted as $t_{\text{pre}}$, is earlier than the spike time of the post-synaptic neuron, denoted as $t_{\text{post}}$, then the weight $w_{ij}$ is increased. This is called long-term potentiation~(LTP). Similarly, when $t_{\text{post}}$ is earlier than $t_{\text{pre}}$, the weight is decreased. This is called long-term depression~(LTD). 

The smaller the time difference, the larger the amount of change in the synaptic weight. This is shown in Fig.~\ref{fig:usual_STDP}. Here, $\Delta t = t_{\text{post}} - t_{\text{pre}}$, and $\Delta w = w_{\text{new}} - w_{\text{old}}$. As shown in the figure, $|\Delta w|$ is maximum when $|\Delta t|=0$, and it is exponentially decreased with larger values of $|\Delta t|$. 
The mathematical equation for the STDP learning rule is 
\begin{equation}
\label{eq:STDP}
\Delta w = \left\{\begin{array}{lc}
        a^{+} \times e^{\displaystyle \big(-\frac{|\Delta t|}{\tau} \big)} &\Delta t > 0 \\
        a^{-} \times e^{\displaystyle \big(-\frac{|\Delta t|}{\tau} \big)} &\Delta t < 0
      \end{array}
    \right.
\end{equation}
where, $a^{+}$ and $a^{-}$ are learning rates, and $\tau$ is the time constant. Note that $a^{+}$ and $a^{-}$ are positive and negative constant values, respectively.

\vskip 2mm
\textit{Analysis of the Learning Rule:} 
The learning rule in \eqref{eq:STDP} creates a form of positive feedback which pushes the synaptic weights towards $+\infty$ or $-\infty$. 
The reason is described in the following. When a synaptic weight $w_{ij}$ is increased, the term $s_i(t)w_{ij}$ in \eqref{eq:LIF} will be larger for the next time, and therefore, the next spike from neuron $i$ is more likely to cause a new spike by neuron $j$. This further increases $w_{ij}$. Since this positive feedback is likely to happen many times, the probability that this weight is pushed towards $+\infty$ is very high.  
Similar situation happens in decreasing the weights, and thus, some of the weights are pushed towards $-\infty$. 
Note that spikes arrive randomly, and hence, during the STDP training, a weight is both increased and decreased many times, depending on the values of $\Delta t$. However, the net effect will be either positive or negative. Therefore, the weights are pushed towards $+\infty$ or $-\infty$.

\vskip 2mm
\textit{Optimized Learning Rule:} 
In computer vision applications, the negative weights are usually clipped to zero and the positive weights to a constant $w_{\text{max}}$. The synaptic weights that are close or equal to $w_{\text{max}}$ form the detected patterns in the input image \cite{Diehl2015a, Iakymchuk2015, Masquelier2007, Kheradpisheh2018, Mozafari2018}. 
In ECG classification, however, the clipping alone is not sufficient, and thus, we also optimize the STDP learning rule as 
\begin{equation}
\label{eq:my_STDP}
\Delta w = \left\{\begin{array}{lc}
        a^{+} \times e^{\displaystyle \big(-\frac{|\Delta t|}{\tau} \big)} \quad \quad &\Delta t > 0 \\
        a^{-} \times \gamma(w) \times e^{\displaystyle \big(-\frac{|\Delta t|}{\tau} \big)} \quad &\Delta t < 0
      \end{array}
    \right.
\end{equation}
where, $\gamma(w)$ is equal to
\begin{equation}
\label{eq:my_STDP_gamma}
\gamma (w) = \frac{1+ w \cdot \gamma_{\text{max}}}{1+w}
\end{equation}

\begin{figure}[tp]
\begin{center}
\centerline{\includegraphics[width=60mm]{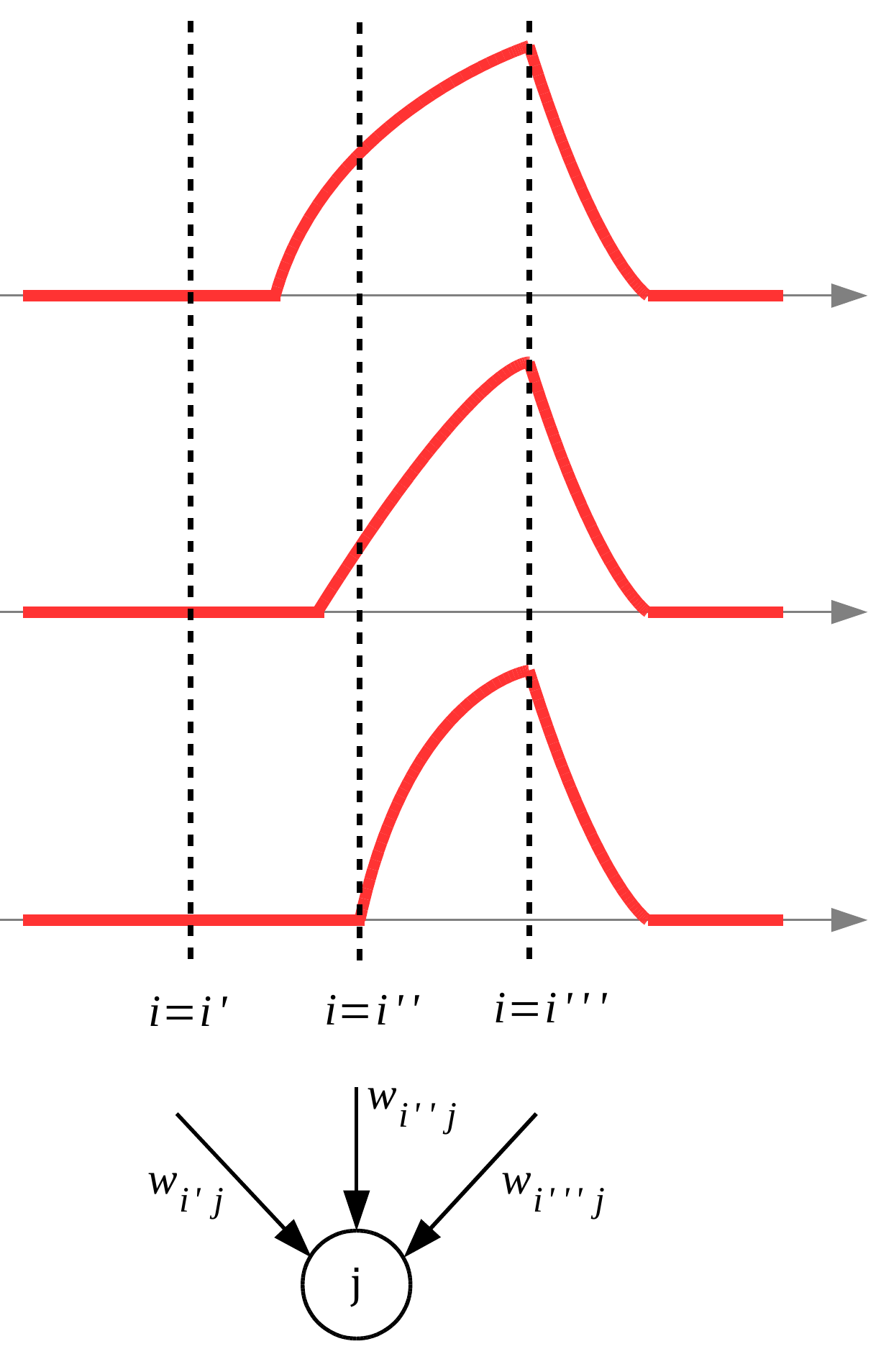}}
\caption{Toy example to demonstrate the intuition behind optimized STDP learning rule in  \eqref{eq:my_STDP}. 
}
\label{fig:non-binary}
\end{center}
\end{figure}

The corresponding curves are shown in Fig.~\ref{fig:our_STDP} and \ref{fig:max_LTD}. The intuition behind this optimization is discussed in the following. Consider the toy example shown in Fig.~\ref{fig:non-binary}, in which three pre-synaptic neurons $i'$, $i''$, and $i'''$ are connected to a post-synaptic neuron $j$. Consider the three input signals shown in Fig.~\ref{fig:non-binary}. Neuron $i'$ generates a very small number of spikes for all the three input signals. This is because the signal amplitude is zero, and hence, the encoder cell which is connected to neuron $i'$ fires at a very low frequency. 
On the other hand, for all the three input signals, neuron $i'''$ generates many spikes. This is because the signal amplitude is large, and hence, the encoder cell which is connected to neuron $i'''$ fires at a high frequency. Since the generated spikes cause neuron $j$ to fire as well, the synaptic weight $w_{i'''j}$ is increased many times. 
This also decreases $w_{i'j}$ because the spike frequency of $i'$ is much lower, and $\Delta t$ is negative in some cases. 
Therefore, in both the learning rules \eqref{eq:STDP} and \eqref{eq:my_STDP}, $w_{i'j}$ and $w_{i'''j}$  are trained as $0$ and $w_{\text{max}}$, respectively. Basically, neuron $j$ is trained to be sensitive to a zero amplitude at location $i'$ and a large peak at location $i'''$ in the input ECG signal.

Now, let's study training of the synaptic weight $w_{i''j}$. Note that spike generation by the encoder cells follows a random process. Hence, it is possible that neuron $i''$ fires before $i'''$. In addition, note that the weights are initialized to random values. Hence, it is possible that $w_{i''j}$ be initialized to a larger value than $w_{i'''j}$. 
Therefore, the learning rule in \eqref{eq:STDP} may train $w_{i''j}$ as $w_{\text{max}}$. It is also possible, depending on the random settings, that $w_{i''j}$ be trained as $0$. 
%OLD:
This random training to either zero or $w_{\text{max}}$ is not desired because it lowers the capability of the network in capturing complex ECG arrhythmia patterns that have various amplitudes in different locations of the input signal.

The proposed learning rule in \eqref{eq:my_STDP}, however, trains $w_{i''j}$ differently. 
Since the three input signals in this example have different amplitudes at location $i''$, we would like $w_{i''j}$ to be trained as somewhere between $0$ and $w_{\text{max}}$. 
%NEW:
This is achieved in the proposed learning rule, because as $w$ increases, $\gamma(w)$ is increased as well, i.e., LTD is amplified (Fig.~\ref{fig:our_STDP}).  Hence, although $w$ is decreased less often than increased, i.e., the likelihood of negative $\Delta t$ is smaller than positive $\Delta t$, the net effect of decreasing $w$ is amplified by making LTD stronger than LTP. As a result, the synaptic weight $w_{i''j}$ is trained as desired.

As shown in Fig.~\ref{fig:max_LTD}, the curve $\gamma(w)$ has a sharp slope near $w=0$. Hence, LTP is amplified quickly with a small increase in $w$. 
This helps to suppress the effect of minor variations in the input ECG signal by lowering the probability that such variations can contribute to the captured patterns. 
This is because it quickly becomes difficult for them to be able to increase the weights. However, once a weight has been increased enough, i.e., a real pattern has been captured, the difficulty stays nearly the same, i.e., the curve $\gamma(w)$ does not increase much because it has a very small slope.

\subsection{Inhibitory Layer}
\label{sec:alg:inhibit}

In order to prevent similar or same patterns from possessing the attention of all neurons in the STDP layer, inhibitory neurons are added to the network as shown in Fig.~\ref{fig:complete_model}.  Inhibitory layer makes the STDP layer more efficient in capturing different patterns \cite{Garner2012}. 

As shown in Fig.~\ref{fig:complete_model}, there is exactly one inhibitory neuron for every neuron in the STDP layer. When a neuron $j$ in the STDP layer fires, its corresponding inhibitory neuron $j$ fires as well and prevents the firing of all the other neurons $j' \neq j$ in the same window in the STDP layer. 
This is achieved through backward synapses with negative weights $w'^q_{jj'}$ which carry the fired spikes from inhibitory neuron $j$ to all neurons $j' \neq j$ in the STDP layer and decrease their membrane potentials. 

The negative synaptic weights $w'^q_{jj'}$ are trained as the following. When neurons from the same window in the STDP layer fire at about the same time, $w'$ is decreased. The inhibitory learning rule is 
\begin{equation}
\label{eq:default_inhibit}
\Delta w' = \left\{\begin{array}{lc}
        b^{-} \quad &|\Delta t| \leq \lambda \\
        b^{+} \quad &|\Delta t| > \lambda
      \end{array}
    \right.
\end{equation}
where, learning rates $b^{+}$ and $b^{-}$ are positive and negative constant values, respectively. When spikes from neurons $j$ and $j'$ in the STDP layer are within $\lambda$ time units from one another, $w'$ is decreased by $b^-$. Otherwise, it is increased by $b^+$. 
Note that $w'$ is clipped to zero, so it always holds a negative value \cite{Srinivasa2013}.

\begin{figure}[tp]
\begin{center}
\centerline{\includegraphics[width=0.7\columnwidth]{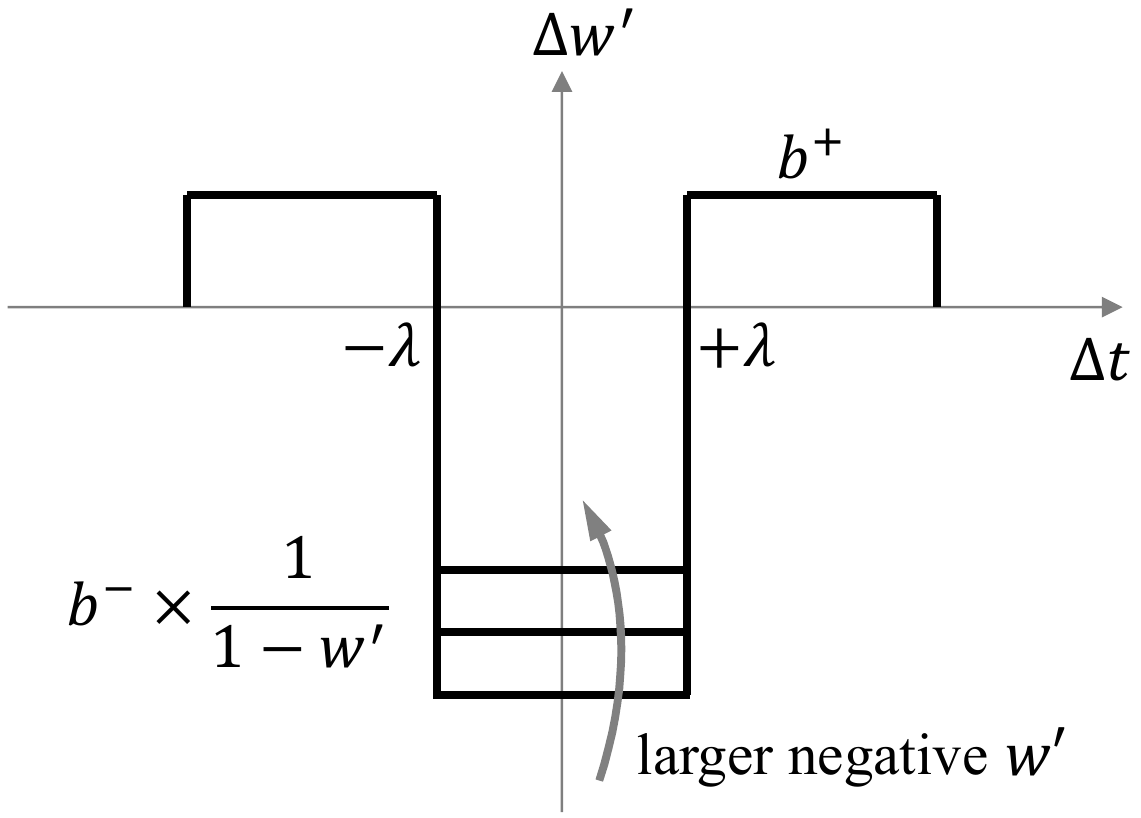}}
\caption{Proposed learning rule for the backward synapses in the inhibitory layer. Note that $w'$ is always kept negative.}
\label{fig:inhibitory}
\end{center}
\end{figure}

\vskip 2mm
\textit{Optimized Learning Rule:} 
When the divergence among the existing patterns are small, the inhibition becomes too strong, which in turn decreases the number of spikes and incapacitates STDP training. To prevent this, we propose to modify the learning rule as shown in \eqref{eq:our_inhibit} and illustrated in Fig.~\ref{fig:inhibitory}. 
\begin{equation}
\label{eq:our_inhibit}
\Delta w' = \left\{\begin{array}{lc}
        b^{-} \times \displaystyle \frac{1}{1-w'} \quad &|\Delta t| \leq \lambda \\
        b^{+} \quad &|\Delta t| > \lambda
      \end{array}
    \right.
\end{equation}

In the proposed learning rule, $\Delta w'$ is a smaller negative value as $w'$ becomes a larger negative value. This modification guarantees that the negative weights do not grow too strong.

\vskip 2mm
\textit{Asymmetric Structure:} 
Our additional strategy for preventing the repercussion of sameness in the detected patterns is to employ an asymmetric structure in the inhibitory layer. In specific, in every epoch, some of the inhibitory neurons are randomly inactivated. This basically distributes the authority among the neurons in the STDP layer. The core idea is inspired by the dropout technique in deep convolutional neural networks \cite{srivastava2014dropout}. 
Note that inhibitory neurons are only present during the training stage, and they are inactivated during the inference (test) stage.

%%%%%%%%%%%%%%%%%%%%%%%%%%%%%%%%%%%%%%%%%%%%%%%%%%%%%%%%%%%%%%%%%%%%%%%%%%%%%%%%
\subsection{Classification via R-STDP}
\label{sec:alg:rstdp}

This layer classifies the patterns which are previously extracted in the STDP layer. Here, reward-modulated STDP (R-STDP) is employed as the classifier. 
R-STDP operates not only based on spike timings but also a modulator signal. 
In the brain, the modulator signal can be a reward or punishment depending on Dopamine secretion \cite{Fremaux2016}. 

As shown in Fig.~\ref{fig:complete_model}, in our proposed solution, R-STDP is used for training synaptic weights $\psi_{jk}^q$ in the final layer. Every neuron in this layer predicts one of the output classes, e.g., the normal class or an arrhythmia class. 
For an input ECG heartbeat, the neuron that fires more often is considered as the winner, i.e., its corresponding class is considered as the predicted class for the input ECG heartbeat. 
Unlike STDP, R-STDP follows a supervised learning procedure. During the training stage, if the prediction of a winner neuron $k$ is correct, a reward signal is applied to all its input synapses. In other words, the reward is applied to all synaptic weights $\psi_{jk}^q$, where $j$ represents all the pre-synaptic neurons from all windows $q$ that are connected to neuron $k$, i.e., the winner neuron. Similarly, if the prediction is incorrect, a punishment signal is applied.

Different learning rules exist for employing R-STDP. Here, we use the following equations which are inspired from the method in \cite{Mozafari2018}. 

\begin{align}
\label{eq:my_RSTDP_reward}
\Delta \wr &=  \left\{\begin{array}{ll}
        a_r^{+} \times \wr(\wr_{\text{max}}-\wr) \quad &t_{\text{post}} > t_{\text{pre}} \\
        a_r^{-} \times \wr(\wr_{\text{max}}-\wr) \quad &t_{\text{post}} \leq t_{\text{pre}}
      \end{array}
    \right. \\
\label{eq:my_RSTDP_punishment}
\Delta \wr &=  \left\{\begin{array}{ll}
        a_p^{-} \times \wr(\wr_{\text{max}}-\wr) &\quad t_{\text{post}} > t_{\text{pre}}\\
        a_p^{+} \times \wr(\wr_{\text{max}}-\wr) &\quad t_{\text{post}} \leq t_{\text{pre}}
           \end{array}
    \right.
\end{align}

In the above equations, $a_r^{+}$ and $a_r^{-}$ are learning rates in the reward situation, while $a_p^{+}$ and $a_p^{-}$ are learning rates in the punishment situation. 
As shown in \eqref{eq:my_RSTDP_reward}, in the reward situation, i.e., when the prediction is correct,  a positive reward proportional to $a_r^{+}$ is given to the synapses whose pre-synaptic neurons spike before the winner neuron, i.e., those with $t_{\text{post}} > t_{\text{pre}}$. A negative reward proportional to $a_r^{-}$ is given to all the other connected synapses, i.e., those with $t_{\text{post}} \leq t_{\text{pre}}$. 
As shown in \eqref{eq:my_RSTDP_punishment}, the negative and positive signals are reversed in the punishment situation, i.e., when the prediction is incorrect.

%NEW:
The term $\wr(\wr_{\text{max}}-\wr)$ highly increases the probability that the trained weights stay near zero or $\wr_{\text{max}}$. This is because $\Delta \wr$ is zero at $\wr=0$ and $\wr=\wr_{\text{max}}$. Basically, once a weight is pushed to either of the two sides, it becomes more difficult to change it. 
Unlike in pattern extraction in the STDP layer, this is desired in classification in the final layer. 
This is because we would like a captured pattern from the STDP layer to have either no or full contribution to the prediction of an output class. 
% 

%%%%%%%%%%%%%%%%%%%%%%%%%%%%%%%%%%%%%%%%%%%%%%%%%%%%%%%%%%%%%%%%%%%%%%%%%%%%%%%%
\subsection{Training Method}
\label{sec:alg:train}

As shown in Fig.~\ref{fig:complete_model}, there are four trainable layers in the proposed model. Since STDP is unsupervised but R-STDP is supervised, the layers are trained successively. 
In specific, first the Gaussian layer is trained. Next, the STDP and inhibitory layers are trained simultaneously, and finally, the R-STDP layer is trained.

All weights in the entire network need to be positive, and therefore, negative weights are clipped to zero. However, the backward synapses in the inhibitory layer need to have negative weights, and therefore, only for these synapses, positive weights are clipped to zero.

%% file: Experiment.tex
\section{Experimental Evaluation}
\label{sec:exp}

\subsection{Classification Performance}
\label{sec:exp:setup}

%code
The proposed solution is implemented based on Brian2 which is a Python package for evaluation of spiking neural networks~\cite{Stimberg}. 

%data
The proposed solution is evaluated and compared with previous works using MIT-BIH ECG arrhythmia database. The ECG signals in this database are independently labeled by two or more cardiologists. 
Two groups of ECG signals called DS1 and DS2 are available in this database. DS1 includes representative samples of different ECG signals and artifacts that an arrhythmia detector might encounter in routine clinical practice. DS2 includes complex arrhythmia patterns \cite{MIT-BIH}. 

%train/test
In order to provide thorough and fair comparisons, we employ the exact same data as the previous works. 
In specific, the following patients from the DS2 group are used as test data: 200, 201, 202, 203, 205, 207, 208, 209, 210, 212, 213, 214, 215, 219, 220, 221, 222, 223, 228, 230, 231, 232, 233 and 234. This is the same set of data employed in \cite{Ince2009, Kiranyaz2016, saeed2019}. 
To train the model for each of the above patients, two sets of data are combined, in specific, randomly selected representative heartbeats from all arrhythmia classes in DS1, plus the first five minutes of the patient's own ECG signal. Employing the first five minutes of the patient's own ECG signal is in compliance with AAMI standards \cite{AAMI} and has been used in many previous works \cite{Hu1997, Ince2009, Kiranyaz2016, saeed2019}. Note that the first five minutes are skipped in the test data. 
The result is shown in Table~\ref{tab:power_comparison}. It compares the proposed solution with previous works, in terms of accuracy and network type. 

\subsection{Real-time Operation and Energy Consumption} 
\label{sec:exp:energy}

The neurons' operating time step is $1$~ms, i.e., the operating frequency is $1$~KHz. Every heartbeat signal is provided to the network for $200$~ms. At the end of this time duration, the network provides its classification result. This configuration supports up to $60~\times~1000~/~200~=~300$ beats per minute, and therefore, it is suitable for real-time operation. 

Once the proposed model is trained, it is simulated according to the method presented in \cite{Cao2015,Qiao2016} in order to estimate its energy consumption in classifying the ECG signals. 
The input ECG signals are applied to the network and the neurons are allowed to fire. When a neuron fires a spike, we consider that $50$~pJ (pico Joules) of energy is consumed \cite{Qiao2016}. The connected output synapses transfer this spike to other neurons. For every spike transfer, i.e., every synaptic event, $147$~pJ of energy is counted \cite{Qiao2016}. 
The simulation results show that the energy consumption is $1.78$~$\mu$J (micro Joules) per beat. This is significantly smaller than previous neural network based solutions.

\begin{table}[pt]
\centering
\caption{Comparing the proposed solution with previous works in terms of accuracy and network type.}
\label{tab:power_comparison}
\renewcommand{\arraystretch}{1.2}
\resizebox{0.8\columnwidth}{!}{
\begin{tabular}{lcc}
\hline
& Accuracy & Type  \\
\hline
Hu \textit{et al.}\cite{Hu1997}                     & 94.8  & MLP          \\
Kachuee \textit{et al.}\cite{Kachuee2018}           & 93.4  & CNN          \\
Kiranyaz \textit{et al.}\cite{Kiranyaz2016}         & 98.6  & FFT, CNN     \\
Saadatnejad \textit{et al.}\cite{saeed2019}         & 99.2  & Wavelet, RNN \\
Ince \textit{et al.}\cite{Ince2009}                 & 97.6  & Wavelet, MLP \\
Kolagasioglu \textit{et al.}\cite{Kolagasioglu2018} & 95.5  & Wavelet, SNN \\
Lee \textit{et al.}\cite{Lee2015}                   & 97.2  & Wavelet      \\
\textbf{Proposed}                                   & 97.9  & SNN          \\ \hline
\end{tabular}}
\end{table}